# Effect of High Energy Heavy Ion Irradiation on c-axis Oriented $MgB_2$ Films

Robert J. Olsson[1,*], Wai-Kwong Kwok[1], Goran Karapetrov[1], Maria Iavarone[1,2], Helmut Claus[1,3], Chad Peterson[+] and George W. Crabtree[1]

[1]Materials Science Division, Argonne National Laboratory, Argonne, IL 60439, USA

[2]INFM–Dipartimento di Scienze Fisiche of the University of Naples "Federico II," Piazzale Tecchio 80, 80125 Naples, Italy

[3]Department of Physics, University of Illinois at Chicago, Chicago, IL 60607-7059, USA

W. N. Kang, Hyeong-Jin Kim, Eun-Mi Choi and Sung-Ik Lee

National Creative Research Initiative Center for Superconductivity and Department of Physics, Pohang University of Science and Technology, Pohang 790-784, Republic of Korea

Abstract

We report on the transport, magnetization, and scanning tunneling spectroscopy measurements on c-axis oriented thin films of $MgB_2$ irradiated with high energy heavy ions of uranium and gold. We find a slight shift in the irreversibility and upper critical field lines to higher temperatures after irradiation. In addition, we observe an increase in the critical current at high temperatures near $T_{c2}$ and only a small change at low temperatures. Furthermore, we find no evidence for the existence of anisotropic pinning induced by heavy ion irradiation in this material. Tunneling spectra in an irradiated sample show a double gap structure with a flat background and very low zero-bias conductance, behaving in much the same way as the pristine unirradiated sample.

Recently, the metallic compound $MgB_2$ was discovered to be superconducting at above T=35K [1]. The distinctive band structure and selective electron-phonon coupling in this material raise many questions about the possibility of high transition temperatures in related compounds. Likewise, in terms of future technological applications, it is important to investigate the behavior of vortex pinning in $MgB_2$. There already exists evidence that grain boundaries do not significantly affect supercurrent flow in this material [2-4], but may arrest the motion of vortices [5]. Questions such as the existence of a vortex liquid state that may reduce pinning close to the upper critical temperature, $T_{c2}$, or whether pinning can be enhanced artificially through irradiation as in the high-$T_c$ superconductors remain to be explored.

In this paper, we investigate the effect of high-energy heavy ion irradiation on the vortex behavior of c-axis oriented thin films of $MgB_2$. Amorphous columnar tracts created by heavy ion irradiation have been demonstrated to be one of the most effective means of enhancing pinning and reducing the vortex liquid state by shifting the irreversibility line to higher temperatures in the cuprate high temperature superconductors [6-8]. In $MgB_2$, we find only a slight enhancement of the critical current at high magnetic fields near $T_{c2}$ and virtually no change in the critical current at low temperatures and at low fields. On the other hand, there is encouraging indication of an increase in the critical current at high magnetic fields near $T_{c2}$, possibly related to the enhancement of the upper critical field after irradiation. However, no indication of anisotropic pinning typically associated with the formation of columnar defects is observed. The irreversibility and upper critical field lines for H ∥ c are shifted slightly to higher temperatures after irradiation, whereas the irreversibility line for H ∥ ab remains virtually unchanged.

The 4000Å thick c-axis oriented $MgB_2$ films used in this study were grown via a pulsed laser deposition technique on an $Al_2O_3$ substrate resulting in an oriented film with the crystallographic c-axis perpendicular to the substrate surface [9]. The film was cut into two pieces; one piece was irradiated with 1.2 GeV $U^{57+}$ ions at the Argonne Tandem Linear Accelerator System (ATLAS) and the other piece was kept as a reference. The ion beam was directed parallel to the c-axis of the film and the irradiation dose was 2 x $10^{11}$ ions/cm$^2$, corresponding to a dose matching field $B_\Phi$=4T, a defect concentration value equivalent to the number of vortices at H=4T with a spacing of about 230Å between defects. A second film was irradiated with 1.4 GeV $Au^{32+}$ ions to a dose matching field of $B_\Phi$=3T. Resistivity was measured using the standard four-probe method. Gold contacts were evaporated onto the plane of the film and gold wires were attached with silver epoxy. For transport measurements, the sample was placed in the bore of two orthogonal superconducting magnets, an 8T solenoid and a transverse 1.5T split-coil, which allowed rotation of the magnetic field vector to investigate anisotropic vortex pinning. The rotation plane of the magnetic field vector was always kept perpendicular to the applied current, preserving the constant Lorentz force configuration on the vortices.

Figure 1 shows the superconducting transition in zero field for the reference and the 1.2GeV uranium irradiated film. The zero field transition temperature, $T_{co}$, is defined by the peak in the temperature derivative of the resistivity (inset Fig. 1) and yields $T_{co}$=38.15K and 38.13K for the unirradiated and irradiated film, respectively. The residual resistivity ratio for the unirradiated film is R(300K)/R(40K) =2.2±0.02, in good agreement with other measurements on c-axis oriented films [10, 11]. The residual resistivity ratio of the irradiated film decreases slightly to 2.04 ± 0.02.

The temperature dependence of the normalized resistivity in the presence of a magnetic field up to H=8Tesla for the unirradiated and irradiated sample is shown in Figure 2. The top panel shows the data for H ∥ c. At high magnetic fields, the resistivity for the irradiated sample (circles) goes to zero at a higher temperature than the unirradiated sample (triangles). However at low fields, this difference decreases rapidly. In contrast, the superconductive resistive transitions for H ∥ ab in the irradiated sample show virtually no change compared to the unirradiated sample.

One of the unique features of the vortex phase diagram in the copper-oxide high temperature superconductor is the existence of a vortex liquid phase, usually indicated by a broad resistive transition in the presence of a magnetic field [12]. The liquid phase is generally marked by ohmic behavior and transforms abruptly into non-ohmic behavior at lower temperatures upon freezing of the vortices into a lattice or glass state. We observe a similar ohmic to non-ohmic behavior in our c-axis oriented $MgB_2$ films. The squares in Figure 2a mark the onset of non-ohmic behavior for the irradiated and unirradiated sample at H=8T. Linear resistivity is observed above this temperature and non-linear resistivity is observed below it. We determined the temperature of the onset of non-ohmic behavior from voltage-current measurements as shown in Figure 3. At high temperatures above T~15K, the curves display a power law behavior with a unit exponent. With decreasing temperature, the exponent deviates from unity. For example, the curve for T= 15.02K, deviates from the dashed line which represents an exponent of unity, marking the temperature at which non-ohmic behavior is first observed.

We determined the critical current $J_c$ from both magnetization measurements using a vibrating sample magnetometer and from transport voltage-current measurements using a 10μV/cm criterion. A typical magnetization curve taken at T=10K is shown in Figure 4. We observe very little change in the size of the magnetization loop for the irradiated and unirradiated samples. The slight difference in the magnetization loop size lies within the sample volume calculation error for the two samples. Critical current from magnetization data was obtained using the Bean critical state model. Figure 5 delineates the temperature dependence of the critical current for the irradiated (open circles) and reference (open triangles) sample. The points joined by the dashed lines are obtained from magnetization data while the points joined by solid lines are obtained from voltage-current measurements close to $T_{c2}$. There is hardly any change in the critical current obtained from magnetization measurements for fields below 6T. Likewise, transport $J_c$ measurements show an enhancement of the critical current with irradiation only at fields above 1 Tesla and near $T_{c2}(H)$ which decreases with increasing field. The observed increase in the critical current at higher fields and lower temperatures may be due to the enhancement of pinning energy at lower temperatures coupled with the increased sampling of the pinning sites with larger number of vortices. In addition, matching of the temperature dependent superconducting coherence length $\xi(T)$ with the defect size may also play a role. Increased values of the irreversibility line and the critical current have been reported in 100μm thick large powder fragments of $MgB_2$ irradiated with protons [5] in which the protons did not exit the sample but were implanted within the material.

Even higher values of the irreversibility line have been reported on unirradiated thin films [14]. Moreover, it has been argued that the grain boundaries in $MgB_2$ films could act as pinning centers for vortices with pinning strength superseding that of the defects created by proton irradiation [5]. However, the random crystal orientation in polycrystalline bulk samples and films make it difficult to draw any clear conclusions. Our use of c-axis oriented $MgB_2$ films over polycrystalline bulk samples and films eliminates the issue of anisotropy from confounding the comparison of vortex behavior in irradiated and unirradiated samples.

In order to directly verify the existence of anisotropic pinning which is one of the hallmarks of columnar defects [15], we measured the angular dependence of the resistivity at a fixed value of the magnetic field and at various temperatures. Figure 6 shows the angular dependence of the resistivity for the irradiated (open circles) and unirradiated (open triangles) samples. Surprisingly, we find no evidence of anisotropic pinning, which if present, should appear as a sharp minima at $\theta=0^o$ [16]. Instead, the angular dependence of the resistivity seems to follow the superconducting anisotropy with no discernible deviation. The absence of anisotropic pinning suggests that columnar defects may not have been created with the high-energy heavy ion irradiation, explaining the weak enhancement of the critical current after irradiation. From TRIM [17] Monte-Carlo calculations for 1.2 GeV uranium ion irradiation on $MgB_2$, a threshold electronic energy loss per collision of dE/dx~25keV/nm was obtained. However, this threshold estimate may be too low for a number of reasons. It is likely that the metallic compound $MgB_2$ may have a significantly higher threshold for track formation. For example, a threshold of ~40kEV/nm was observed in $NiZr_2$ and $Ni_3B$ samples with still higher thresholds obtained for more symmetric structures [18]. In cuprate superconductors, it has been reported that strings of isolated "droplet" or cascade defected regions first form at lower energies, with continuous tracks forming at higher energies of irradiation [19, 20]. These cascade defects usually occur along the columnar defect axis, although at low enough energies, they may become disordered [19]. The existence of such "droplets" may explain the small enhancement in pinning observed for H || c and the isotropic nature of the pinning, but does not explain the lack of any change in the irreversibility line for H || ab. It is also likely that the thermal spike model [21] which has been used to explain the formation of columnar defects in both elements and alloys may not be applicable here as recently reported in $(U,T)Be_{13}$ [22].

Scanning tunneling microscopy on a second $MgB_2$ film irradiated with 1.4GeV gold ions was performed as yet another direct check for the existence of columnar defects. The STM used in our measurements is a home built system operating at 4.2K in helium exchange gas. The tip was made from Pt-Ir wire, either mechanically sharpened or electrochemically etched. A topographical scan at 4.2K over a region of 3500Å x 3500Å found no evidence of any columnar defects. The dose matching field was $B_\Phi$=3T for this film, corresponding to a columnar defect separation of ~260Å, placing it well within the range of the scanned area.

Current-voltage characteristics (I-V) and conductance spectra (dI/dV vs V) were recorded at different locations of the scanning area. The differential conductance dI/dV vs V curves were recorded using a standard lock-in technique with a small ac modulation superimposed to a slowly varying bias voltage while the feedback loop was interrupted.

The amplitude of the ac modulation was fixed at 0.2 mV-0.4 mV, below the intrinsic thermal broadening at 4.2K.

The conductance spectra obtained in films without any surface treatment are very broad and similar to those already reported earlier in $MgB_2$ pellets [23]. In an alternative procedure, the film was etched for 50 seconds in bromine (Br 1% in pure ethanol), rinsed in pure ethanol and dried in $N_2$ gas. After this treatment the sample was mounted on the STM stage in helium exchange gas and very quickly cooled down to 4.2 K. No columnar defects were observed from scanning the topography of the etched sample. Typical I-V and dI/dV vs T spectra recorded on the sample surface at junction resistance of 0.1G . are shown in Figure 7. The conductance spectra, normalized at the conductance value at –20mV, reveal a c-axis tunneling gap structure, a flat background and a very low zero-bias conductance consistent with only a little smearing other than thermal broadening. The peak shows up at 2.9 meV with a weak shoulder at 6.5 meV, symmetrically for both injection and emission of electrons. These spectra are absolutely reproducible with location and show tunneling resistance in the range of 0.1-1 GΩ. Other tunneling spectroscopy experiments [7,8], on pellets and powder, have indicated the presence of a double gap consistent with the theoretical prediction of Liu et al [24] for two-gap superconductivity in $MgB_2$ in the clean limit. According to ref. [24] the small and large gaps should arise from the 3D and 2D sheets of the Fermi surface, respectively. This scenario seems to be supported by specific heat measurements [25], low temperature Raman scattering experiments [26] and photoemission experiments [27]. In case of c-axis oriented films, the contribution from the 3D Fermi surface should dominate the tunneling conductance as we observe here.

The tunneling conductance spectra taken at different magnetic fields perpendicular to the film surface and at different locations show the spatially averaged pair-breaking effect of the magnetic field (see Fig. 8). The magnetic field dramatically increases the number of quasiparticle states in the gap and smears the superconducting peaks. No additional features in the gap were observed in applied field.

We conclude from STM measurements that the irradiated sample shows no vestige of any columnar defects and instead behaves in much the same way as a pristine unirradiated sample. However, we cannot rule out the possibility of sub-surface point defects incurred from heavy ion irradiation.

Our results are summarized in the vortex phase diagram of Figure 9. The figure depicts an upward shift in the H || c irreversibility line of the irradiated film compared with the reference sample for H > 1T. The irreversibility lines, $H_{irr}$, were obtained from the onset temperature of non-ohmic behavior from the V-I curves of Figure 3. Also shown are the upper critical field lines for H || c and H || ab of the unirradiated and irradiated samples defined for each field as the temperature where the resistivity first starts to decrease from the normal state value. This is obtained from the *onset* of the peak in the temperature derivative of the resistivity (dρ/dT), defined as the temperature where the value for dρ/dT first extends beyond the background scatter of the normal state resistivity. Typically at the onset temperature, the resistance has fallen only 3-5% below the normal state value. The value of the anisotropic ratio $\gamma=H_{c2}(||ab)/H_{c2}(||c)=2.0 \pm 0.2$ is in general agreement with both c-axis oriented films [10, 11] and aligned particles [28] reported elsewhere. The difference between the irreversibility line and the upper critical field line suggests the existence of an observable vortex liquid regime in $MgB_2$. We find

a noticeable increase in $H_{c2}(T) \parallel c$ for the irradiated film whereas virtually no change is observed for $H_{c2} \parallel ab$. Consequently, the anisotropy reduces to $\gamma=1.6 \pm 0.2$ after irradiation. The enhancement of $H_{c2} \parallel c$ may be caused by a reduction in the mean free path, $l$, due to enhanced carrier scattering after irradiation since the coherence length $\xi \propto (\xi_o l)^{1/2}$ and $H_{c2} \propto \xi^{-2}$.

In conclusion, we present one of the first studies of high-energy heavy ion irradiation in a c-axis oriented film of $MgB_2$. We observe a shift in the irreversibility line and the upper critical field line to higher temperatures after irradiation. An enhancement of the critical current was observed at high fields near $T_{c2}$. However, magnetization measurements only see a very weak enhancement in the critical current at low temperatures. We find no evidence for anisotropic pinning in the irradiated film that would indicate the formation of columnar defects. This is corroborated with STM measurements which find no vestige of any columnar defects on the film's surface. The STM tunneling spectra show a double gap structure, flat background and very low zero-bias conductance, behaving in much the same way as a pristine unirradiated sample. Finally, the difference between the irreversibility line and the upper critical field suggests the existence of a vortex liquid regime in this material, similar to that found in the high $T_c$ copper oxide superconductors.

**Acknowledgement**

This work was supported by the U.S. Department of Energy, Basic Energy Sciences-Material Sciences under contract No. W-31-109-ENG-38 (RJO,WKK, GK, MI, HC, CP, GWC) and by the Ministry of Science and Technology of Korea through the Creative

Research Initiative Program (WNK, HJK, EMC, SIL). M.I. would like to thank the INFM for financial support. C.P. would like to thank the U.S. Department of Energy-ERULF program at Argonne National Laboratory for financial support.

**Figure captions**

**Figure 1.** Temperature dependence of the resistivity in zero applied field for the unirradiated and irradiated c-axis oriented $MgB_2$ films. Inset shows the temperature derivative of the superconductive resistive transition used to determine $T_{co}$.

**Figure 2.** (a) Resistivity versus temperature for applied magnetic fields H=0, 0.5, 1, 2, 4, 6 and 8 Tesla along the c-axis of the irradiated (open circles) and unirradiated (open triangles) films. The large open square delineates the onset of non-ohmic behavior obtained from voltage-current measurements. (b) Resistivity versus temperature for H=0, 0.5, 1, 2, 3, 4, 5, 6, 7 and 8 Tesla along the ab plane of the film perpendicular to the current direction.

**Figure 3.** Logarithmic voltage – current measurements at various temperatures for H=8T || c. The dashed line corresponds to a power law behavior with a power of unity. Note the deviation from this behavior beginning with the T=15.02K curve, signaling the onset of non-ohmic behavior.

**Figure 4.** Typical magnetization curve of the irradiated and unirradiated film taken at T=10K

**Figure 5.** Critical current versus temperature obtained from V-I data (connected lines) and from magnetization data (dashed lines) at several magnetic fields for the irradiated (open circles) and the unirradiated (open triangles) films.

**Figure 6.** Angular dependence of the resistivity at various temperatures for the irradiated (open circles) and unirradiated (open triangles) films at H=0.5T.

**Figure 7.** I-V and the corresponding dI/dV at the junction resistance 0.1 G$\Omega$.

**Figure 8.** Field dependence of the conductance spectra.

**Figure 9.** Vortex phase diagram for the irradiated and unirradiated films depicting the irreversibility lines for H || c and the upper critical field lines for H || c and H || ab.

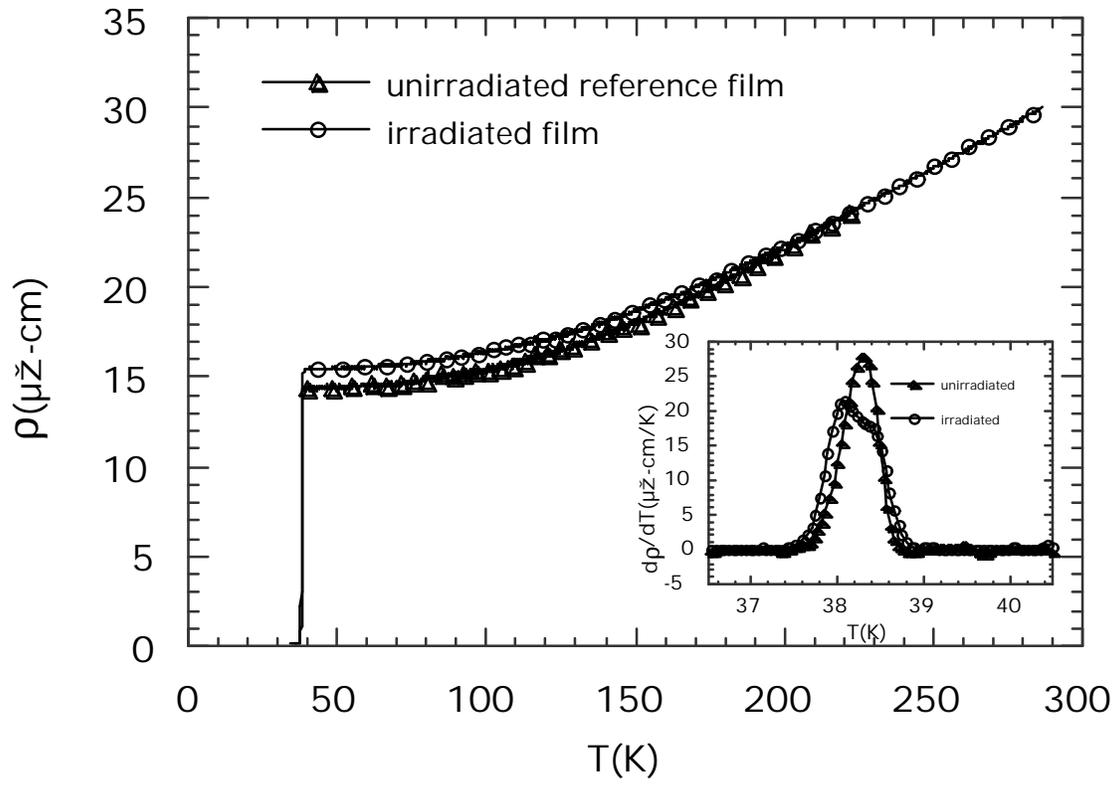

Figure 1

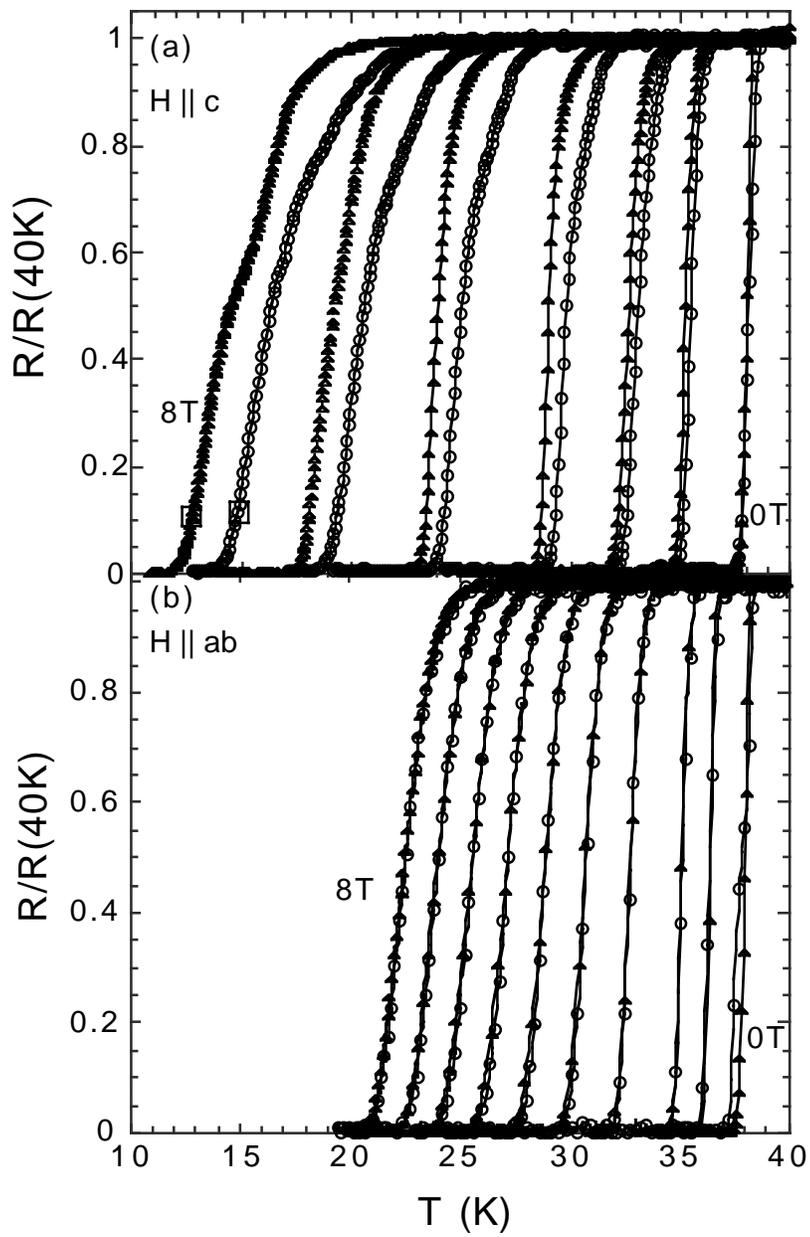

FIGURE 2

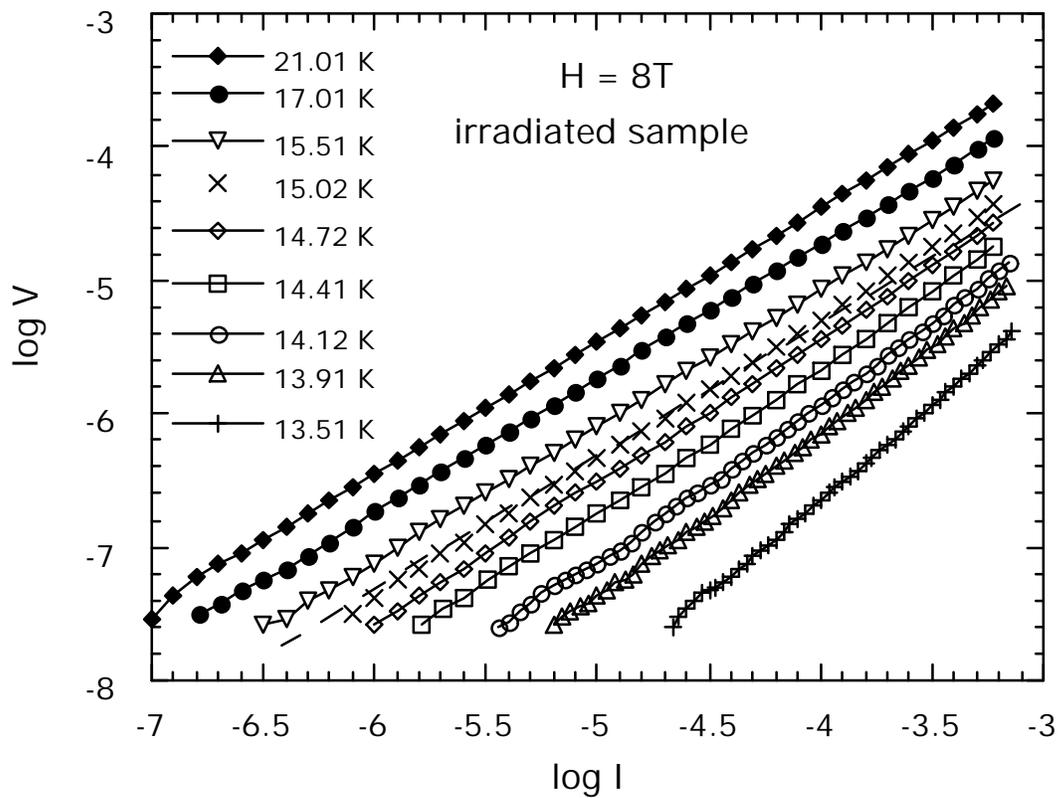

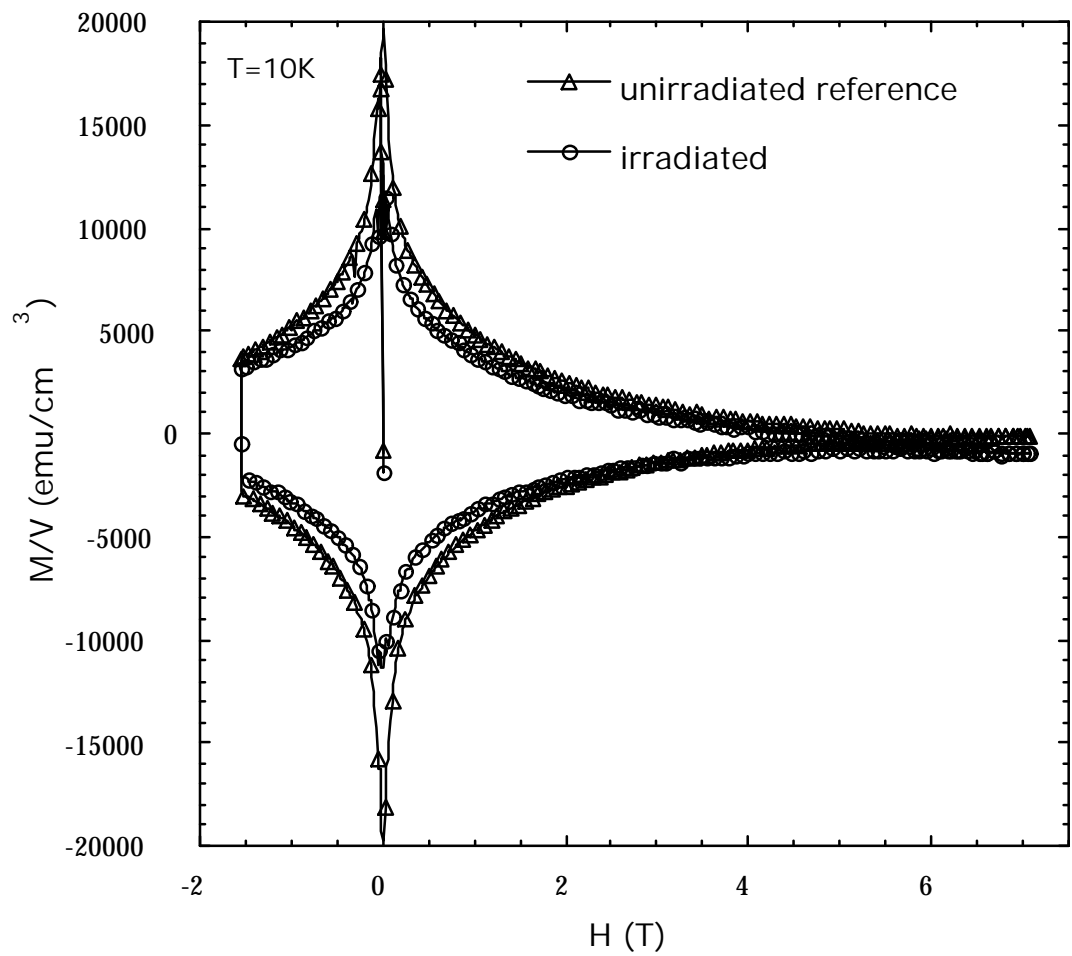

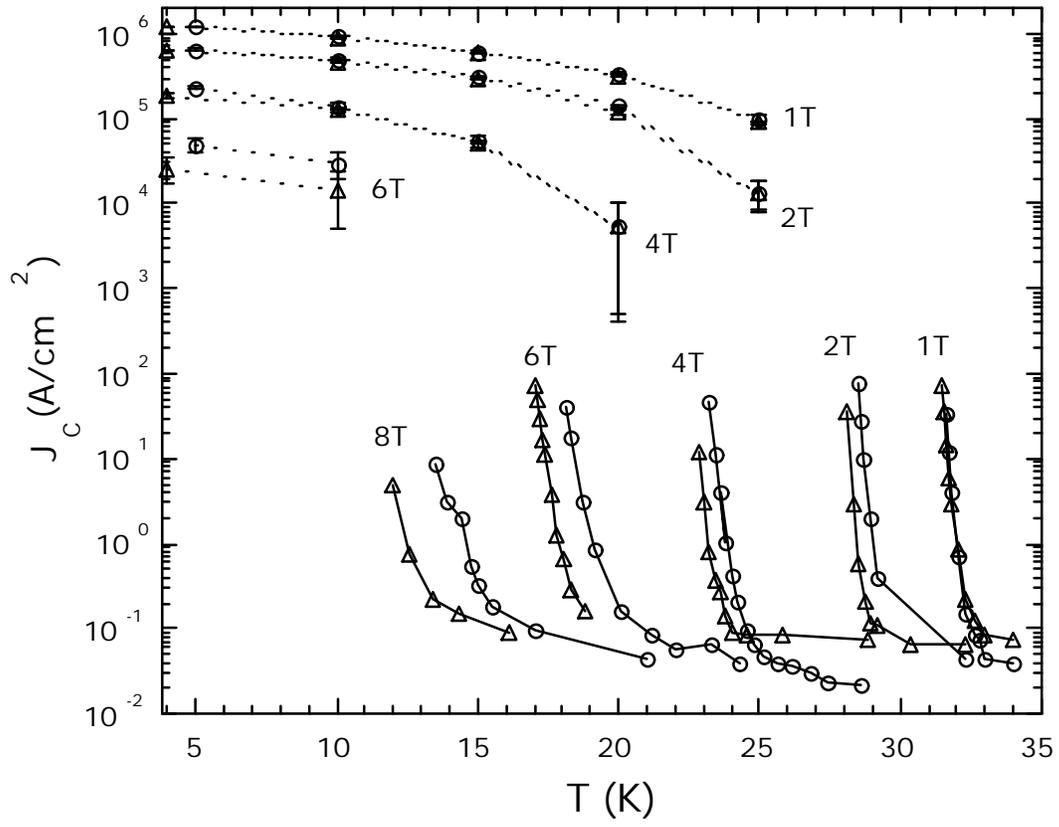

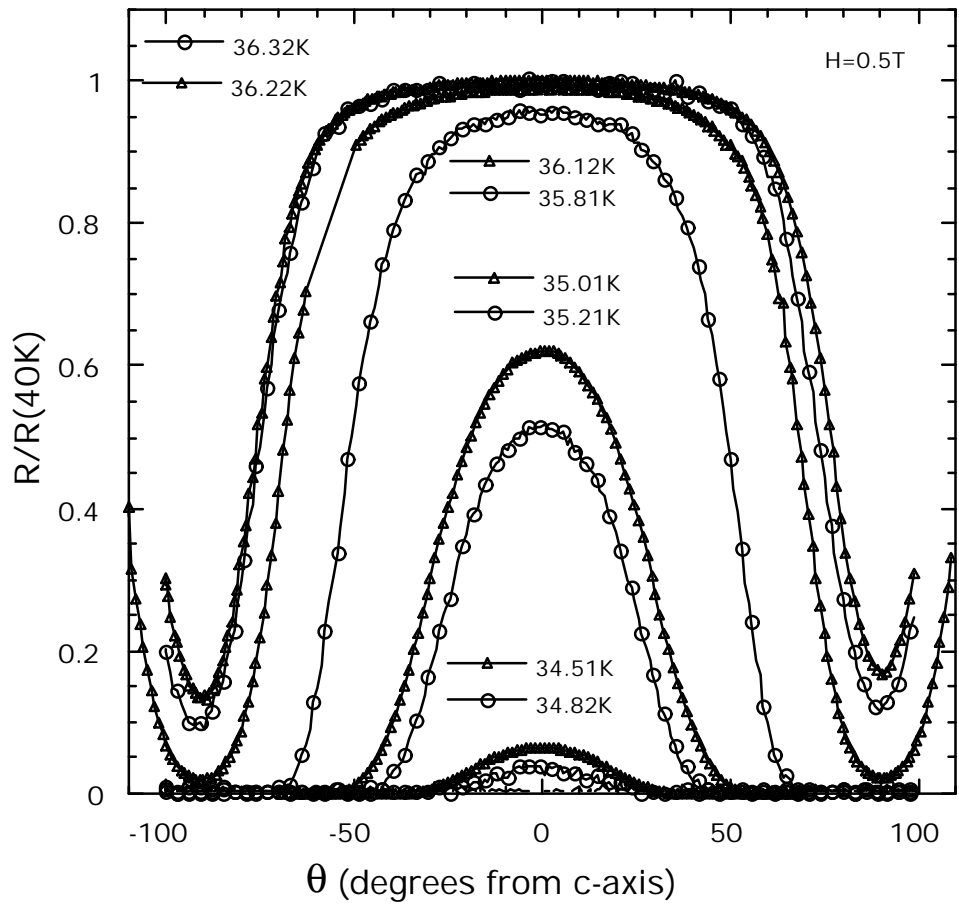

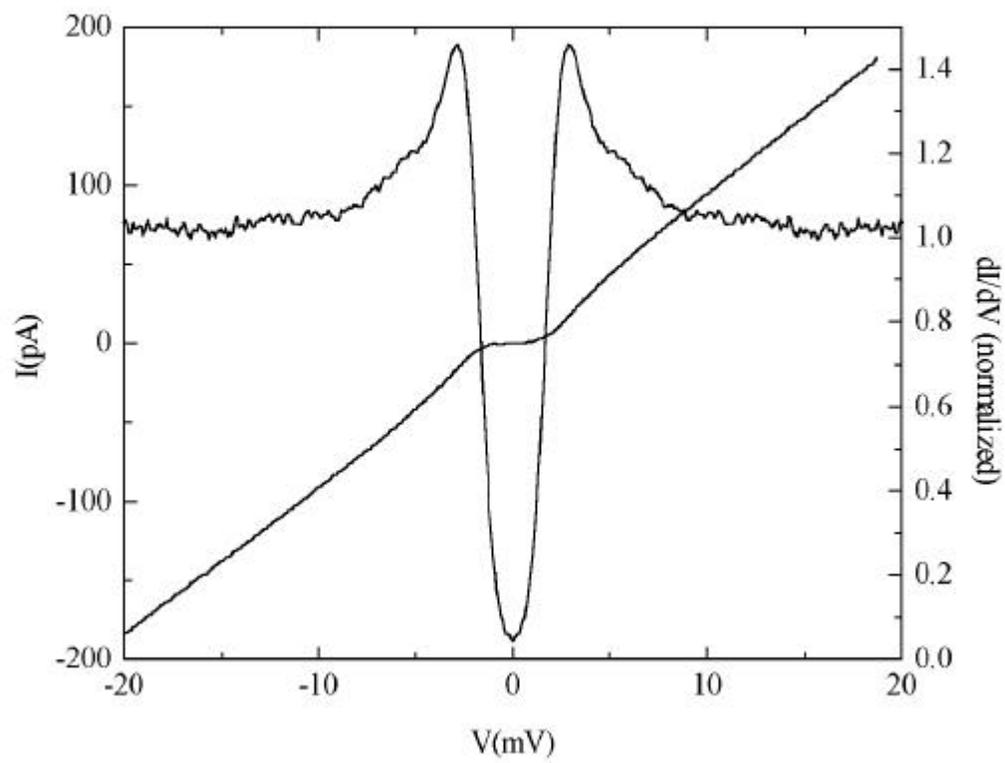

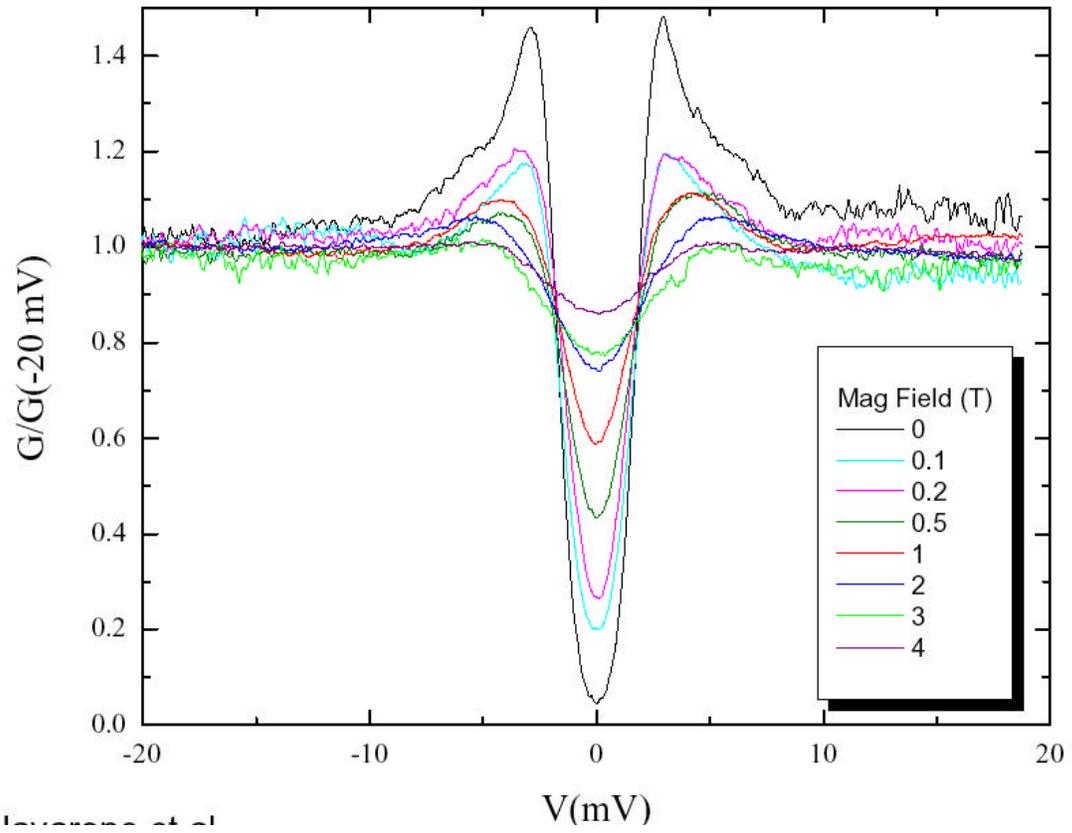

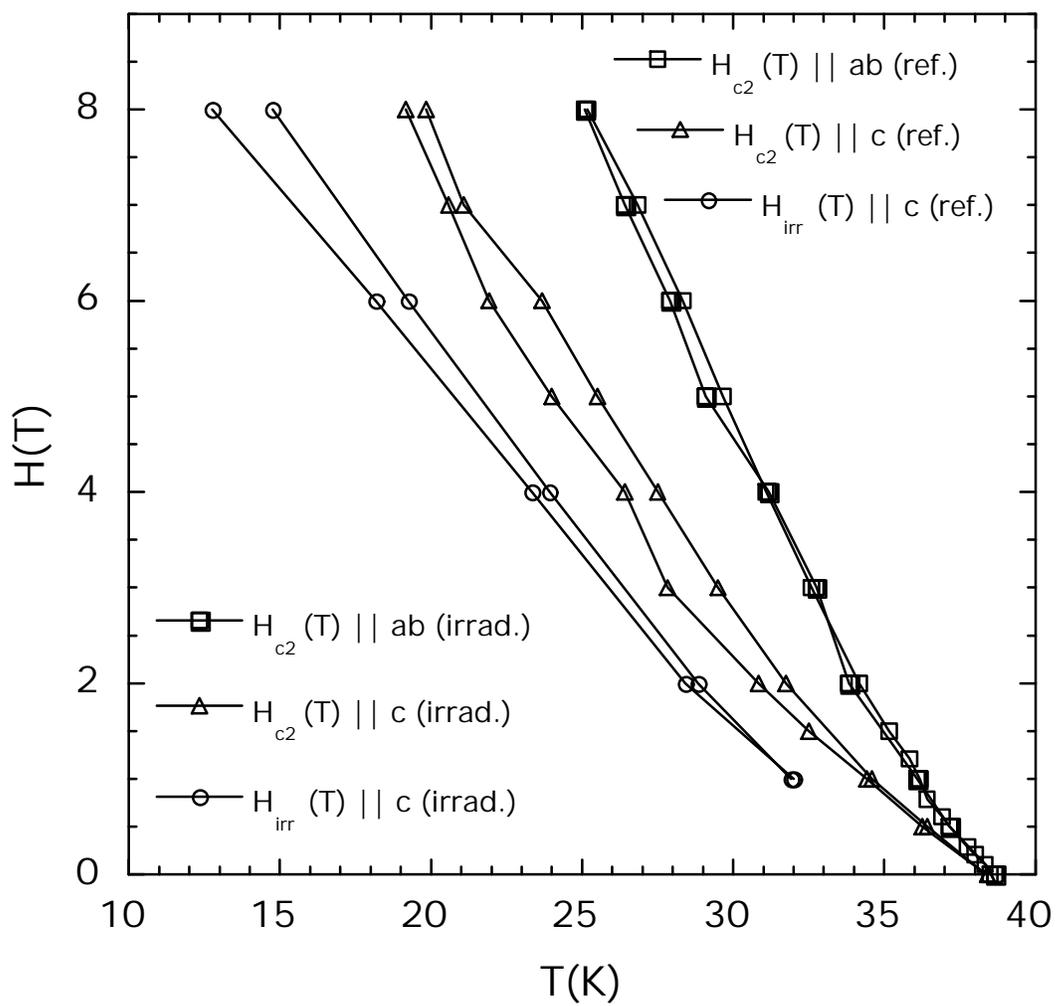